\pdfoutput=1

\documentclass[aps,superscriptaddress,twocolumn,showpacs,showkeys,amsmath,amssymb,nofootinbib,tightenlines,floatfix]{revtex4}

\usepackage{graphicx}
\usepackage{dcolumn}
\usepackage{epstopdf} 

\newcommand{\ldbrace}{\bigl \{ \negmedspace \bigl \{}
\newcommand{\rdbrace}{\bigr \} \negmedspace \bigr \}}
\newcommand{\Ldbrace}{\Bigl \{ \negmedspace \Bigl \{}
\newcommand{\Rdbrace}{\Bigr \} \negmedspace \Bigr \}}
\newcommand{\Ldbbrace}{\Biggl \{ \negmedspace \Biggl \{}
\newcommand{\Rdbbrace}{\Biggr \} \negmedspace \Biggr \}}

\allowdisplaybreaks[4]
\begin{document}

\title{Combinatorial Entropy for Distinguishable Entities in Indistinguishable States}


\author{Robert K. Niven}
\email{r.niven@adfa.edu.au}
\affiliation{School of Aerospace, Civil and Mechanical Engineering, The University of New South Wales at ADFA, Northcott Drive, Canberra, ACT, 2600, Australia.}
\altaffiliation{Presently at Niels Bohr Institute, University of Copenhagen, Copenhagen \O, Denmark.}



\begin{abstract}
 \noindent The combinatorial basis of entropy by Boltzmann can be written $H= {N}^{-1} \ln \mathbb{W}$, where $H$ is the dimensionless entropy of a system, per unit entity, $N$ is the number of entities and $\mathbb{W}$ is the number of ways in which a given realization of the system can occur, known as its statistical weight. Maximizing the entropy (``MaxEnt'') of a system, subject to its constraints, is then equivalent to choosing its most probable (``MaxProb'') realization. For a system of distinguishable entities and states, $\mathbb{W}$ is given by the multinomial weight, and $H$ asymptotically approaches the Shannon entropy. In general, however, $\mathbb{W}$ need not be multinomial, leading to different entropy measures.  
\\
\indent This work examines the allocation of distinguishable entities to non-degenerate or equally degenerate, {indistinguishable} states. The non-degenerate form converges to the Shannon entropy in some circumstances, whilst the degenerate case gives a new entropy measure, a function of a multinomial coefficient, coding parameters, and Stirling numbers of the second kind. 
\end{abstract}

\maketitle


\section{Introduction}

Of the many interpretations of the entropy concept, the {\it combinatorial} (or {\it probabilistic}) basis of entropy was given by Boltzmann \cite{Boltzmann_1877} and Planck \cite{Planck_1901} in the famous equation: 
\begin{equation}
S_N  = NS = k \ln \mathbb{W}  
\label{eq:Boltz1}
\end{equation}
where $S_N$ is the total thermodynamic entropy of the system, $S$ is the entropy per unit entity, $N$ is the number of entities, $\mathbb{W}$ is number of ways in which a specified realization of a system can occur, known as its statistical weight, and $k$ is the Boltzmann constant. This can be rewritten to give the dimensionless entropy \cite{Vincze_1972, Grendar_G_2001, Niven_CIT, Niven_2007}:
\begin{equation}
H = \frac{S}{k} = \frac{1}{N} \ln \mathbb{W}
\label{eq:Boltz2}
\end{equation}
If the weight is of multinomial form, i.e.\ $\mathbb{W}_{mult} = N! / \prod\nolimits_{i=1}^s n_i! =  \bigl (  \begin{smallmatrix} N \\  {n_1, n_2, \dots, n_s}  \\  \end{smallmatrix} \bigr )$, where $n_i$ is the number of entities in the $i$th state, from $s$ such states, then in the asymptotic limits $N \to \infty$, $n_i \to \infty$, $\forall i$, using the Stirling approximation \cite{Stirling_1730}, $\ln m! \approx m \ln m -m$ (or using Sanov's theorem \cite{Sanov_1957}), the entropy converges to the Shannon function \cite{Shannon_1948}:
\begin{equation}
H_{Sh}  =  - \sum\limits_{i = 1}^s {p_i \ln p_i } 
\label{eq:Shannon}
\end{equation}
where $p_i=n_i/N$ is the probability of the $i$th state. However, it must be recognised that a system may not be of multinomial weight.   
The best-known examples are the three distributions examined in quantum physics \cite{Bose_1924, Einstein_1924, Einstein_1925, Fermi_1926, Dirac_1926}:  
\begin{list}{$\bullet$}{\topsep 2pt \itemsep 2pt \parsep 0pt \leftmargin 8pt \rightmargin 0pt \listparindent 0pt
\itemindent 0pt}
\item {\it Degenerate Maxwell-Boltzmann statistics}, in which distinguishable entities are allocated to distinguishable states, with $g_i$ degenerate sub-states within each state (this reduces to the multinomial case for $g_i=1,\forall i$); 
\item {\it Bose-Einstein statistics}, in which indistinguishable entities are allocated to distinguishable, degenerate states; and
\item {\it Fermi-Dirac statistics}, also with indistinguishable entities allocated to distinguishable, degenerate states, but with a maximum of one entity per state;
\end{list}
The weights and entropy functions of these statistics are well known \citep[e.g.][]{Tolman_1938, Davidson_1962, Niven_2005}.  In such cases, maximisation of the combinatorial entropy defined by \eqref{eq:Boltz2} (``MaxEnt''), subject to the constraints on a system, {\it always} yields the realization of maximum probability (``MaxProb'') (or, in the non-asymptotic case, a distribution close to the maximum) \cite{Vincze_1972, Grendar_G_2001, Niven_CIT, Niven_2007}. This provides a much stronger (purely probabilistic) definition of the entropy concept than that given by axiomatic or information-theoretic reasoning.

The aims of this work are (i) to review the concept of distinguishability, so often used in physics (\S\ref{Disting}), and (ii) to derive the statistical weight and entropy of a sys\-tem in which distinguishable entities (balls) are allocated to {\it indistinguishable} states (boxes), for both non-degenerate and equally degenerate cases (\S\ref{Non-Degen}-\ref{Degen}). Although this occupancy problem has a long history \cite{Stirling_1730, Abramowitz_S_1965, Comtet_1974, Char_2002} and is included 
in combinatorial classification schemes \cite{Johnson_K_1977, Fang_1985, Zwillinger_2003}, its connection to entropy does not appear to have been examined previously. 

\section{\label{Disting}On Distinguishability}

The concept of {\it distinguishability} strongly affects the choice of statistic used for analysis. Firstly, the entities and/or states of a system might be fundamentally indistinguishable (as is currently believed in quantum physics); the statistic is thus pre-ordained.  A second, more interesting case is when we {\it choose} whether to distinguish the entities and/or states, based on the {\it purpose} for which the entropy measure will be used.  This is illustrated by the allocation of physicists (entities) to the seats of a bus (boxes).  Four scenarios arise:
\begin{list}{$\bullet$}{\topsep 2pt \itemsep 2pt \parsep 0pt \leftmargin 8pt \rightmargin 0pt \listparindent 0pt
\itemindent 0pt}
\item A conference organiser is requested by physicists A, B and C for window seats, while X and Y require seats near the door; also, everyone is concerned about the likely argument between Q and T, should they be seated together.  In this case, it is necessary to distinguish both the physicists and seats, leading to degenerate multinomial (Maxwell-Boltzmann) statistics (or a variant thereof, with a maximum of $m$ physicists per seat).
\item The bus company wishes to model the wear and tear on its seats.  Here they have no interest in distinguishing the physicists, but need to distinguish the seats.  This leads to Bose-Einstein statistics (or an intermediate variant).
\item Alternatively, the conference organiser does not have any seat-specific requests, but is concerned about who will sit together. Here the physicists are distinguishable but the seats are not, leading to a new type of statistic (examined herein).
\item Finally, a more disinterested observer (e.g.\ a traffic engineer) does not care who the passengers are, or where they sit, but needs to model whether the bus schedule is sufficient to meet demand. Here both the physicists and seats are indistinguishable.
\end{list}
Such considerations lead naturally to the ``subjective'' (or ``observer-dependent'') view of the entropy concept, a viewpoint vigorously defended by Jaynes \cite{Jaynes_1957} \citep[c.f.][]{Vincze_1972, Niven_2007}.  

\section{\label{Non-Degen}The Non-Degenerate Case}

We now consider the number of ways in which $N$ distinguishable balls can be allocated to $s$ non-degenerate, indistinguishable boxes, to give the realization $\{n_i\}$ of numbers of balls in each box (the boxes being unlabelled), as shown in Figure 1. This statistical weight can be denoted 
$\mathbb{W}_{D:I} = \ldbrace  \begin{smallmatrix} N \\  {n_1, n_2, \dots, n_s}  \\  \end{smallmatrix} \rdbrace $, 
with $\sum\nolimits_{i=1}^s n_i=N$.  It is known \cite{Jordan_1947} that the number of ways to arrange $N$ distinguishable balls in $k$ non-empty indistinguishable boxes (for $k \le s$) is given by the Stirling number of the second kind $\bigl \{  \begin{smallmatrix} N \\  k  \\  \end{smallmatrix} \bigr \}$, the first few values of which are listed in Table \ref{tab:Stirling2}. These satisfy the recurrence relation \cite{Jordan_1947}:
\begin{equation}
\bigl \{  \begin{smallmatrix} N \\  k  \\  \end{smallmatrix} \bigr \} = \bigl \{  \begin{smallmatrix} {N-1} \\  {k-1}  \\  \end{smallmatrix} \bigr \} + k \bigl \{  \begin{smallmatrix} {N-1} \\  k  \\  \end{smallmatrix} \bigr \}, \quad \bigl \{  \begin{smallmatrix} N \\  1  \\  \end{smallmatrix} \bigr \} = \bigl \{  \begin{smallmatrix} N \\  N  \\  \end{smallmatrix} \bigr \} =1.
\end{equation}
By combinatorial enumeration, it is readily determined that $\bigl \{  \begin{smallmatrix} N \\  k  \\  \end{smallmatrix} \bigr \}$  does not give $\mathbb{W}_{D:I}$; e.g.\ $\ldbrace  \begin{smallmatrix} 5 \\  {3, 1, 1}  \\  \end{smallmatrix} \rdbrace= 10$ and $\ldbrace  \begin{smallmatrix} 5 \\  {2, 2, 1}  \\  \end{smallmatrix} \rdbrace= 15$; it is their sum which gives the Stirling number $\bigl \{  \begin{smallmatrix} 5 \\  3  \\  \end{smallmatrix} \bigr \}=25$.  By definition, this gives the general result:
\begin{equation}
\bigl \{  \begin{smallmatrix} N \\  s  \\  \end{smallmatrix} \bigr \}  = \bigl \{  \begin{smallmatrix} N \\  k  \\  \end{smallmatrix} \bigr \} = \sum\limits_{\text{all } \{n_i\} \text{, fixed } k}  \ldbrace  \begin{smallmatrix} N \\  {n_1, n_2, \dots, n_k, 0, \dots, 0}  \\  \end{smallmatrix} \rdbrace
\label{eq:thm1}
\end{equation}
in which the zeroes (unfilled states) extend from $n_{k+1}$ to $n_s$. Of course, some Stirling numbers $\bigl \{  \begin{smallmatrix} N \\  k  \\  \end{smallmatrix} \bigr \}$ permit only one realization, e.g.:
\begin{gather}
\bigl \{  \begin{smallmatrix} N \\  1  \\  \end{smallmatrix} \bigr \} =  \ldbrace  \begin{smallmatrix} N \\  {N, 0, \dots, 0}  \\  \end{smallmatrix} \rdbrace = 1
\label{eq:thm1_eg1} \\
\bigl \{  \begin{smallmatrix} N \\  N  \\  \end{smallmatrix} \bigr \} =  \Ldbrace  \begin{smallmatrix}  \\ \phantom{|} \\ N \\  {\underbrace {\scriptstyle 1,1,\dots,1}_{\scriptstyle N{\text{ times}}},0,\dots,0} \\  \end{smallmatrix} \Rdbrace = 1
\label{eq:thm1_eg2} \\
\bigl \{  \begin{smallmatrix} N \\  N-1  \\  \end{smallmatrix} \bigr \} =  \Ldbrace  \begin{smallmatrix}  \\ \phantom{|} \\ N \\  {\underbrace {\scriptstyle 2,1,\dots,1}_{\scriptstyle N-1},0,\dots,0} \\  \end{smallmatrix} \Rdbrace = \bigl (  \begin{smallmatrix} N \\  2  \\  \end{smallmatrix} \bigr )
\label{eq:thm1_eg3}
\end{gather} 

\begin{figure}[t]
\setlength{\unitlength}{0.6pt}
  \begin{picture}(280,230)
   \put(0,0){\includegraphics[width=55mm]{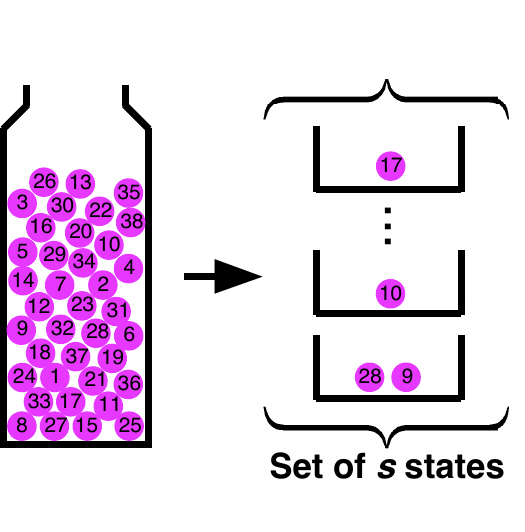} }
   \end{picture}
\caption{Allocation of distinguishable balls to indistinguishable boxes.}
\label{fig:DtoI}
\end{figure}

\begin{table}[t]
\begin{tabular}{c c p{15pt} p{15pt} p{15pt} p{15pt} p{15pt} l l l} 
\hline
&	{\bf k=1}	&{\bf 2}	&{\bf 3}	&{\bf 4}	&{\bf 5}	&{\bf 6}	&{\bf 7}\\
{\bf N=1}	&1							\\
{\bf 2}	&1	&1						\\
{\bf 3}	&1	&3	&1					\\
{\bf 4}	&1	&7	&6	&1	&	&	&	&$\dots$\\
{\bf 5}	&1	&15	&25	&10	&1			\\
{\bf 6}	&1	&31	&90	&65	&15	&1		\\
{\bf 7}	&1	&63	&301	 &350	&140	 &21	&1	\\
	&	&	&$\vdots$	 \\
\hline
\end{tabular}
\caption[Stirling numbers of the second kind.]{Stirling numbers of the second kind $\bigl \{  \begin{smallmatrix} N \\  k  \\  \end{smallmatrix} \bigr \}$.}
\label{tab:Stirling2}
\end{table} 

What can we say about $\mathbb{W}_{D:I}$?  Firstly, it is unaffected by any zeroes amongst the $n_i$, since we could arbitrarily add unfilled states to \eqref{eq:thm1}, without change. Secondly, as the states are unlabelled, it is meaningless to permute the $n_i$; e.g.\ $\ldbrace  \begin{smallmatrix} 5 \\  {2, 2, 1}  \\  \end{smallmatrix} \rdbrace $ and $\ldbrace  \begin{smallmatrix} 5 \\  {2, 1, 2}  \\  \end{smallmatrix} \rdbrace$ refer to the {\it same} rea\-lization $\{2,2,1\}$.  This is quite different to the multinomial weight; e.g. $\bigl (  \begin{smallmatrix} 5 \\  {2, 2, 1}  \\  \end{smallmatrix} \bigr )$ and $\bigl (  \begin{smallmatrix} 5 \\  {2, 1, 2}  \\  \end{smallmatrix} \bigr )$ are num\-erically equal, but represent different realizations $[2,2,1]$ and $[2,1,2]$. The $D:I$ statistic thus has fewer realizations than the multinomial case.

It can in fact be shown that:
\begin{equation}
\begin{split}
\mathbb{W}_{D:I} &= \ldbrace  \begin{smallmatrix} N \\  {n_1, n_2, \dots, n_k, 0, \dots, 0}  \\  \end{smallmatrix} \rdbrace = \biggl (  \begin{matrix} N \\  {n_1, n_2, \dots, n_k}  \\  \end{matrix} \biggr ) \Bigl( \prod\limits_{j=1}^N \frac {1} {r_j!} \Bigr) \\
&= \frac {N!} { \Bigl( \prod\limits_{i=1}^k  {n_i!} \Bigr) \Bigl( \prod\limits_{j=1}^N {r_j!} \Bigr) }
= \frac {N!} { \Bigl( \prod\limits_{i=1}^s  {n_i!} \Bigr) \Bigl( \prod\limits_{j=1}^N {r_j!} \Bigr) }
\end{split}
\label{eq:W_DtoI}
\end{equation}
where $r_j \ge 0$ is the number of occurrences of integer $j$ in the set $\{ n_i \}$, or its {\it repetitivity} (note that the zeros are not counted), whence $\sum\nolimits_{j=1}^{N} r_j =k$.  Proof of \eqref{eq:W_DtoI} considers the successive filling of boxes:\ $\mathbb{W}_{D:I}$ must equal the number of ways to choose $n_1$ balls from $N$ balls, multiplied by the number of ways to choose $n_2$ balls from $N-n_1$ balls, and so on, for $k$ boxes; the product must then be divided by the number of ways that each multiply-occurring integer $j$ can occur in the set $\{n_i\}$, given by $r_j!$, to account for the indistinguishable boxes.  For example,  
$\ldbrace  \begin{smallmatrix} 5 \\  {2, 2, 1}  \\  \end{smallmatrix} \rdbrace= \bigl(  \begin{smallmatrix} 5 \\  2  \\  \end{smallmatrix} \bigr) \bigl(  \begin{smallmatrix} 3 \\  2  \\  \end{smallmatrix} \bigr) \bigl(  \begin{smallmatrix} 1 \\  1  \\  \end{smallmatrix} \bigr) \tfrac{1}{2!} = 15$, or, since the order of filling is immaterial, $\ldbrace  \begin{smallmatrix} 5 \\  {2, 2, 1}  \\  \end{smallmatrix} \rdbrace= \bigl(  \begin{smallmatrix} 5 \\  1  \\  \end{smallmatrix} \bigr) \bigl(  \begin{smallmatrix} 4 \\  2  \\  \end{smallmatrix} \bigr) \bigl(  \begin{smallmatrix} 2 \\  2  \\  \end{smallmatrix} \bigr) \tfrac{1}{2!} = 15$.  Generalising this result:
\begin{align}
\begin{split}
&\mathbb{W}_{D:I} = \\
&\biggl (  \begin{matrix} N \\  n_1 \\  \end{matrix} \biggr ) \biggl (  \begin{matrix} N-n_1 \\  n_2 \\  \end{matrix} \biggr ) \dots \biggl (  \begin{matrix} N-n_1- \dots - n_{k-1} \\  n_k \\  \end{matrix} \biggr ) 
\Bigl( \prod\limits_{j=1}^N \frac {1} {r_j!} \Bigr) 
\end{split}
\label{eq:W_DtoI2}
\end{align}
The product of binomial coefficients in \eqref{eq:W_DtoI2} is simply the multinomial coefficient, giving \eqref{eq:W_DtoI}; the last form in \eqref{eq:W_DtoI} can be used when $k$ is not known in advance. $\square$ 

The weight has been given previously \cite{Abramowitz_S_1965, Comtet_1974, Char_2002} in the form:
\begin{equation}
\mathbb{W}_{D:I} = \bigl(N; r_1, r_2, \dots, r_N)^{\prime}
= \frac {N!} { \prod\limits_{j=1}^N \hspace{2pt}  (j!)^{r_j} \hspace{2pt} {r_j!}  }
\label{eq:W_DtoI_Abram}
\end{equation}
Recognising each $j$ term in \eqref{eq:W_DtoI_Abram} as one of the $n_j$ terms, this reduces to \eqref{eq:W_DtoI}. It is, however, much less useful for the derivation of an entropy function. 

Also needed is the sum of the weights, given by an {\it incomplete Bell number} \cite{Char_2002}:
\begin{align}
\begin{split}
B(N,s) &= \sum\limits_{k=1}^s \Bigl \{  \begin{matrix} N \\  k  \\  \end{matrix} \Bigr \} \\
&= \sum\limits_{k=1}^s \hspace{4pt} \sum\limits_{\begin{smallmatrix}  \text{all } \{n_i\} \\ \text{fixed } k \end{smallmatrix} }  \Ldbrace  \begin{matrix} N \\  {n_1, n_2, \dots, n_k, 0, \dots, 0}  \\  \end{matrix} \Rdbrace
\end{split}
\label{eq:Bell}
\end{align}
This reduces to the usual Bell number $B_N$  \cite{Zwillinger_2003} for $s=N$.

The non-asymptotic entropy, denoted $\#$, can now be calculated using \eqref{eq:Boltz2} and \eqref{eq:W_DtoI}:
\begin{align}
\begin{split}
H_{D:I}^\# &= \frac{1}{N} \ln \mathbb{W}_{D:I} \\
&= \frac{1}{N} \sum\limits_{i=1}^s \Bigl( \frac{n_i}{N} \ln N! - \ln n_i! \Bigr) - \frac{1}{N} \sum\limits_{j=1}^{N} \ln r_j!
\end{split}
\label{eq:H_DtoI_exact}
\end{align}
where the $\ln N!$ term is brought inside the first sum using $\sum\nolimits_{i=1}^s n_i=N$. Application of the ``traditional'' asymptotic limits $N \to \infty$, $n_i \to \infty$, $\forall i$, using the Stirling approximation, as well as the corresponding limits $r_{j \ne \infty} = 0$, $r_{\infty} = k$ then gives, for $k \nrightarrow \infty$:
\begin{equation}
H_{D:I} =- \sum\limits_{i = 1}^s {p_i \ln p_i }  -  \lim\limits_{N \to \infty} \Bigl( \frac{1}{N} \ln k! \Bigr) = - \sum\limits_{i = 1}^s {p_i \ln p_i }
\label{eq:H_DtoI_limit}
\end{equation}
$H_{D:I}$ therefore converges to the Shannon entropy \eqref{eq:Shannon} in the traditional limits. However, this is not the full picture, as shown by the following examples. 

\vspace{6pt}
\noindent {\bf Examples}: 
The most probable realization(s) for both multinomial and $D:I$ statistics, determined by enumerating all realizations, are listed for several values of $N$ in Tables \ref{tab:mult}-\ref{tab:DtoI}, for two situations: (i) $N=s$, and (ii) $s=3$.  To summarise:
\begin{list}{$\bullet$}{\topsep 2pt \itemsep 2pt \parsep 0pt \leftmargin 12pt \rightmargin 0pt \listparindent 0pt
\itemindent 0pt}
\item As expected, the most probable realization of the multinomial statistic,  subject only to the natural constraint, corresponds in all cases to the uniform distribution $[N/s,\dots,N/s]$, or, if unable to achieve this (due to quantisation of the balls or boxes), to a set of equiprobable local maxima centred on the uniform distribution. 
\item In contrast, the $D:I$ statistic for $N=s$ gives a step function (``staircase'') with many unfilled boxes. The number of filled states $k$, the number of steps and the width of each step all increase with $N$, but the number of unfilled states $(s-k)$ increases even more rapidly. This preferential bunching (``cohesion'') of the filled states is very different to their preferential spreading in the multinomial case. In consequence, applying the ``traditional'' asymptotic limit $n_i \to \infty, \forall i$ (in addition to $N \to \infty$) is inappropriate for $k<s$, since it does not account for the stepped nature of the distribution (with $n_i \ll \infty$ in many states), and also distorts the role of the $r_j$. Thus for $k<s$, the asymptotic limit in \eqref{eq:H_DtoI_limit} does not apply.  On the other hand, for $s=3$ (an example of $N \gg s$), the most probable realizations are again non-uniform, but all boxes are filled (except for $N \le 3$).  As evident from the table, the filling will become more uniform as $N \to \infty$, which will be consistent with $n_i \to \infty, \forall i$; this then gives the asymptotic limit of \eqref{eq:H_DtoI_limit}. 
\end{list}
The $D:I$ statistic thus has very different convergence properties to the multinomial, being more strongly dependent on small values of $n_i$; in the limit $N \to \infty$, it asymptotically approaches the Shannon entropy for $N \gg s$ and $k=s$.  Outside of these bounds, more detailed analysis is needed to identify any asymptotic limits; until then, the non-asymptotic entropy $H_{D:I}^{\#}$ \eqref{eq:H_DtoI_exact} must be used. 

\begin{table*}[p]
\begin{tabular}{p{15pt} p{15pt} p{210pt} p{40pt} p{50pt} l }  
\hline
$N$	&$s$	&MaxProb realization $[n_i]$	&Maxima	&$\mathbb{W}$ (each)	&$\mathbb{P}=\frac{\mathbb{W}}{\sum \mathbb{W}}$ (each) \\
\hline
1	&1	&[1]	&1	&1	&1\\
2	&2	&[1, 1]	&1	&2	&0.5\\
3	&3	&[1, 1, 1]	&1	&6	&0.222222\\
4	&4	&[1, 1, 1, 1]	&1	&24	&0.093750\\
5	&5	&[1, 1, 1, 1, 1]	&1	&120	&0.038400\\
10	&10	&[1, 1, 1, 1, 1, 1, 1, 1, 1, 1]	&1	&3.63E+06	&3.629E-04\\
20	&20	&[1, \dots, 1]	&1	&2.43E+18	&2.320E-08\\
30	&30	&[1, \dots, 1]	&1	&2.65E+32	&1.288E-12\\
40	&40	&[1, \dots, 1]	&1	&8.16E+47	&6.749E-17\\
50	&50	&[1, \dots, 1]	&1	&3.04E+64	&3.424E-21\\
\hline
1	&3	&[1, 0, 0], [0, 1, 0], [0, 0, 1]	&3	&1	&0.333333\\
2	&3	&[1, 1, 0], [1, 0, 1], [0, 1, 1]	&3	&2	&0.222222\\
3	&3	&[1, 1, 1]	&1	&6	&0.222222\\
4	&3	&[1, 1, 2], [1, 2, 1], [2, 1, 1]	&3	&12	&0.148148\\
5	&3	&[1, 2, 2], [2, 1, 2], [2, 2, 1]	&3	&30	&0.123457\\
10	&3	&[3, 3, 4], [3, 4, 3], [4, 3, 3]	&3	&4200	&0.071127\\
20	&3	&[6, 7, 7], [7, 6, 7], [7, 7, 6]	&3	&1.33E+08	&0.038151\\
30	&3	&[10, 10, 10]	&1	&5.55E+12	&0.026961\\
40	&3	&[13, 13, 14], [13, 14, 13], [14, 13, 13]	&3	&2.41E+17	&0.019853\\
50	&3	&[16, 17, 17], [17, 16, 17], [17, 17, 16]	&3	&1.15E+22	&0.016005\\
\hline
\end{tabular}
\caption []{Most probable realizations for the multinomial statistic, using $\mathbb{W}=\mathbb{W}_{mult}$ and $\sum \mathbb{W} = s^N$.}
\label{tab:mult}
\end{table*} 
\begin{table*}[p]
\begin{tabular}{p{15pt} p{15pt} p{210pt} p{40pt} p{50pt} l }  
\hline
$N$	&$s$	&MaxProb realization $\{n_i\}$	&Maxima	&$\mathbb{W}$ (each)	&$\mathbb{P}=\frac{\mathbb{W}}{\sum \mathbb{W}}$ (each)\\
\hline
1	&1	&\{1\}	&1	&1	&1\\
2	&2	&\{1, 1\}, \{2, 0\}	&2	&1	&0.5\\
3	&3	&\{2, 1, 0\}	&1	&3	&0.6\\
4	&4	&\{2, 1, 1, 0\}	&1	&6	&0.4\\
5	&5	&\{2, 2, 1, 0, 0\}	&1	&15	&0.288462\\
10	&10	&\hspace{-5pt}$\begin{array}[t]{l}\{3, 2, 2, 1, 1, 1, 0, 0, 0, 0\}, \{3, 2, 2, 2, 1, 0, 0, 0, 0, 0\},\\ \{3, 3, 2, 1, 1, 0, 0, 0, 0, 0\}, \{4, 3, 2, 1, 0, 0, 0, 0, 0, 0\} \end{array}$	&4	&12600	&0.108644\\
20	&20	&\hspace{-5pt}$\begin{array}[t]{l}\{4, 3, 3, 2, 2, 2, 2, 1, 1, 0, \dots, 0\},\\ \{4, 4, 3, 3, 2, 2, 1, 1, 0, \dots, 0\} \end{array}$	&2	&1.83E+12	&0.035443\\
30	&30	&\{5, 4, 4, 3, 3, 3, 2, 2, 2, 1, 1, 0, \dots, 0\}	&1	&1.54E+22	&0.018214\\
40	&40	&\{5, 4, 4, 4, 3, 3, 3, 3, 2, 2, 2, 2, 1, 1, 1, 0, \dots, 0\}	&1	&1.14E+33	&0.007265\\
50	&50	&\{6, 5, 5, 4, 4, 4, 3, 3, 3, 3, 2, 2, 2, 2, 1, 1, 0, \dots, 0\}	&1	&7.40E+44	&0.003986\\
\hline
1	&3	&\{1, 0, 0\}	&1	&1	&1\\
2	&3	&\{1, 1, 0\}, \{2, 0, 0\}	&2	&1	&0.5\\
3	&3	&\{2, 1, 0\}	&1	&3	&0.6\\
4	&3	&\{2, 1, 1\}	&1	&6	&0.428571\\
5	&3	&\{2, 2, 1\}	&1	&15	&0.365854\\
10	&3	&\{5, 3, 2\}	&1	&2520	&0.256046\\
20	&3	&\{8, 7, 5\}	&1	&9.98E+07	&0.171680\\
30	&3	&\{11, 10, 9\}	&1	&5.05E+12	&0.147059\\
40	&3	&\{15, 13, 12\}	&1	&2.09E+17	&0.103236\\
50	&3	&\{18, 17, 15\}	&1	&1.02E+22	&0.085360\\
\hline
\end{tabular}
\caption[]{Most probable realizations for the D:I statistic, using $\mathbb{W}=\mathbb{W}_{D:I}$ \eqref{eq:W_DtoI} and $\sum \mathbb{W} = B(N,s)$ \eqref{eq:Bell}.}
\label{tab:DtoI}
\end{table*} 

\section{\label{Degen}The Equally Degenerate Case}

We now consider a simple degenerate form of the $D:I$ statistic, in which each indistinguishable state contains $g$ indistinguishable sub-states, with $n_{im}$ entities in each sub-state, whence $\sum\nolimits_{m=1}^g n_{im} = n_i$. The weight can be denoted:
\begin{equation}
\mathbb{W}_{D:I(g)} = \ldbrace  \begin{smallmatrix} N \\  {n_1, n_2, \dots, n_s}  \\  \end{smallmatrix} \rdbrace _{(g)} = \Ldbbrace  \begin{smallmatrix} & N &\\  n_{11}, &\dots, &n_{s1} \\ {\scriptstyle \vdots} & &\vdots  \\ n_{1g}, &\dots, &n_{sg}  \\  \end{smallmatrix} \Rdbbrace
\end{equation}
Using the reasoning of \S\ref{Non-Degen}, the weight of each realization is given by the weight \eqref{eq:W_DtoI} of filling {\it of} the states, multiplied by the number of ways of filling {\it within} each state; each component of the latter must contain a sum over the possible number of filled sub-states $\gamma=1,...,\min(n_i,g)$. This gives, for fixed $k$: 
\begin{align}
\begin{split}
&\mathbb{W}_{D:I(g)} = \Ldbrace  \begin{matrix} N \\  {n_1, \dots, n_k, 0, \dots, 0}  \\  \end{matrix} \Rdbrace _{(g)}  \\
&= \frac {N!} { \Bigl( \prod\limits_{i=1}^k  {n_i!} \Bigr) \Bigl( \prod\limits_{j=1}^N {r_j!} \Bigr) } 
\\ & \qquad  \times
\prod\limits_{i=1}^k \hspace{4pt} \sum\limits_{\gamma=1}^{\min(g,n_i)} \sum\limits_{\begin{smallmatrix}  \text{all } \{n_{im}\} \\ \text{fixed } \gamma \end{smallmatrix} }  \frac {n_i!} { \Bigl( \prod\limits_{m=1}^{\gamma}  {n_{im}!} \Bigr) \Bigl( \prod\limits_{\ell=1}^{n_i} {r_{i\ell}!} \Bigr) }  
\end{split}
\label{eq:W_DtoIg_long}
\end{align}
where $r_{i \ell}$ is the repetitivity of $\ell$ in the set $\{ n_{im} \}$. Using \eqref{eq:thm1} and \eqref{eq:Bell}, this simplifies to: 
\begin{align}
\begin{split}
\mathbb{W}_{D:I(g)} 
&= \frac {N!} { \Bigl( \prod\limits_{i=1}^k  {n_i!} \Bigr) \Bigl( \prod\limits_{j=1}^N {r_j!} \Bigr) }  \prod\limits_{i=1}^k \hspace{4pt}  \sum\limits_{\gamma=1}^{\min(g,n_i)} \Bigl \{  \begin{matrix} n_i \\  \gamma  \\  \end{matrix} \Bigr \} \\
&= \frac {N!} { \Bigl( \prod\limits_{i=1}^k  {n_i!} \Bigr) \Bigl( \prod\limits_{j=1}^N {r_j!} \Bigr) }  \prod\limits_{i=1}^k \hspace{4pt}  B(n_i,\min(n_i,g)) 
\end{split}
\label{eq:W_DtoIg}
\end{align}
For non-degenerate states $\bigl \{  \begin{smallmatrix} n_i \\  1  \\  \end{smallmatrix} \bigr \}=1, \forall i$, hence \eqref{eq:W_DtoIg} reduces to \eqref{eq:W_DtoI}.  As with the non-degenerate case, the products over $i$ in \eqref{eq:W_DtoIg} can be extended to $s$ instead of $k$, since for unfilled states we can take $\sum\nolimits_{\gamma=1}^{0} \bigl \{  \begin{smallmatrix} 0 \\  \gamma  \\  \end{smallmatrix} \bigr \} = \bigl \{  \begin{smallmatrix} 0 \\  0  \\  \end{smallmatrix} \bigr \} =1$, or alternatively $B(0,0)=1$. 

The non-asymptotic entropy for the simple degenerate case is, from \eqref{eq:Boltz2} and \eqref{eq:W_DtoIg}:
\begin{align}
\begin{split}
&H_{D:I(g)}^\# = \frac{1}{N} \ln \mathbb{W}_{D:I(g)} 
= \frac{1}{N} \sum\limits_{i=1}^s \biggl( \frac{n_i}{N} \ln N!  \\
& - \ln n_i!  + \ln \sum\limits_{\gamma=1}^{\min(g,n_i)} \Bigl \{  \begin{matrix} n_i \\  \gamma  \\  \end{matrix} \Bigr \}  \biggr ) - \frac{1}{N} \sum\limits_{j=1}^{N} \ln r_j! 
\end{split}
\label{eq:H_DtoIg_exact}
\end{align}
In the Stirling-approximate limits $N \to \infty$, $n_i \to \infty, \forall i$, for which $r_{j \ne \infty} = 0$ and $r_{\infty} = k$, we can see (e.g.\ from Table \ref{tab:Stirling2}) that each sum over $\gamma$ in \eqref{eq:H_DtoIg_exact} will be dominated by its largest term $\bigl \{  \begin{smallmatrix} n_i \\  \gamma_i^\#  \\  \end{smallmatrix} \bigr \}$, where $1 \ll \gamma_i^\# \ll n_i$. Applying the Jordan limit $\bigl \{  \begin{smallmatrix} n \\  a  \\  \end{smallmatrix} \bigr \} \approx a^n/a!$ as $n \to \infty$ \cite{Jordan_1947} then gives:
\begin{align}
\begin{split}
H_{D:I(g)} &=- \sum\limits_{i = 1}^s {p_i \ln \frac{p_i}{\gamma_i^\#} }  -  \lim\limits_{N \to \infty}  \frac{1}{N} \Bigl( \ln k! +  \ln \gamma_i^{\#} !   \Bigr)    \\
&=- \sum\limits_{i = 1}^s {p_i \ln \frac{p_i}{\gamma_i^\#} } \qquad \text{for } k \nrightarrow \infty, \gamma_i^\# \nrightarrow \infty
\end{split}
\label{eq:H_DtoIg_limit}
\end{align}
This closely resembles the degenerate Maxwell-Boltzmann entropy $H_{MB}= - \sum\nolimits_{i = 1}^s {p_i \ln ({p_i}/{g_i}) }$, where $g_i$ is the degeneracy of state $i$ \cite{Tolman_1938, Davidson_1962, Niven_2005}. However, as shown in \S\ref{Non-Degen}, this form does not reflect the behaviour of this statistic when unfilled states or sub-states are present, for which $H_{D:I(g)}^\#$ \eqref{eq:H_DtoIg_exact} must be used.

\section{\label{Concl}Conclusions}

The statistical weight for the allocation of distinguishable entities to {indistinguishable} states is derived herein, for both non-degenerate or equally degenerate states. The weight is obtained as a function of a multinomial coefficient, a set of coding parameters, and (for the degenerate case) a set of Stirling numbers of the second kind or of incomplete Bell numbers. Using Boltzmann's combinatorial definition (the ``Boltzmann principle''), the non-asymptotic entropy functions are then obtained.  For fully filled states, the non-degenerate and degenerate entropies converge respectively to the Shannon and degenerate Maxwell-Boltzmann functions, but not otherwise. 

This study illustrates the importance of the combinatorial definition of entropy, for which the maximum entropy position (``MaxEnt'') gives the most-probable (``MaxProb'') realization of the system (or, in the non-asymptotic case, a distribution close to the maximum). For systems which follow the $D:I$ statistic, blind application of MaxEnt based on the Shannon entropy \eqref{eq:Shannon} will give the most probable realization only in special circumstances.  Given the long history of  Bose-Einstein and Fermi-Dirac statistics in physics, for the allocation of indistinguishable entities to distinguishable states \cite{Bose_1924, Einstein_1924, Einstein_1925, Fermi_1926, Dirac_1926, Tolman_1938, Davidson_1962, Niven_2005}, it is surprising that the entropy functions for the opposite occupancy problem do not appear to have been examined previously.

\begin{acknowledgments}
The author thanks the organisers of the CTNEXT07 conference, Catania, Italy, at which this work was presented; the European Commission for financial support by a Marie Curie Fellowship under FP6; and The University of New South Wales for sabbatical leave.
\end{acknowledgments}


\bibliographystyle{aipproc}   

\begin{thebibliography}{99}
\bibitem{Boltzmann_1877} L. Boltzmann, {Wien. Ber.} 76 (1877) 373; English transl.: J. Le Roux (2002) {\it http://www.essi.fr/$\sim$leroux/}.
\bibitem{Planck_1901} M. Planck, {Annalen der Physik} 4 (1901) 553. 
\bibitem{Vincze_1972} I. Vincze, 
{Progress in Statistics (European Meeting of Statisticians, Budapest, Hungary, 1972)}, 2 (1974) 869-895.
\bibitem{Grendar_G_2001} M. Grend\'ar, Jr.\, M. Grend\'ar, What is the question that MaxEnt answers? A probabilistic interpretation, {\it in} A. Mohammad-Djafari (ed.), {Bayesian Inference and Maximum Entropy Methods in Science and Engineering} (MaxEnt 2000), AIP, Melville, 2001, 83-94.
\bibitem{Niven_CIT} R. K. Niven, Combinatorial information theory: I. Philosophical basis of cross-entropy and entropy, {\it arXiv:0512017}, 2005-2007.
\bibitem{Niven_2007} R.K. Niven, Origins of the combinatorial basis of entropy, MaxEnt2007, 8-13 July 2007, Saratonga Springs, NY, in press, {\it arXiv:0708.1861}.
\bibitem{Stirling_1730} J. Stirling, {Methodus Differentialis: Sive Tractatus de Summatione et Interpolatione Serierum Infinitarum}, Gul. Bowyer, London, 1730.
\bibitem{Sanov_1957}I. N. Sanov,
Mat. Sbornik, 42 (1957) 11Ð44 (Russian).
\bibitem{Shannon_1948} C.E. Shannon, {Bell Sys. Tech. J.} 27 (1948) 379; 623.
\bibitem{Bose_1924} S.N. Bose, {Z. Phys.} 26 (1924) 178.
\bibitem{Einstein_1924}A. Einstein, {Sitzungsber. Preuss. Akad. Wiss. Phys. Math. Kl} (1924) 261.
\bibitem{Einstein_1925}A. Einstein, {Sitzungsber. Preuss. Akad. Wiss. Phys. Math. Kl} (1925) 3.
\bibitem{Fermi_1926}E. Fermi, {Z. Phys.} 36 (1926) 902.
\bibitem{Dirac_1926}P.A.M. Dirac, {Proc. Roy. Soc.} 112 (1926) 661.
\bibitem{Tolman_1938} R.C. Tolman, {The Principles of Statistical Mechanics}, Oxford Univ. Press, London, 1938.
\bibitem{Davidson_1962} N. Davidson, {Statistical Mechanics}, McGraw-Hill, NY, 1962.
\bibitem{Niven_2005}R.K. Niven, {Phys. Lett. A} 342(4) (2005) 286.
\bibitem{Abramowitz_S_1965}M. Abramowitz, I.A. Stegun,  Handbook of Mathematical Functions with Formulas, Graphs and Mathematical Tables, U.S. Govt Printing Office, Washington D.C, 1965.
\bibitem{Comtet_1974}L. Comtet, Advanced Combinatorics, rev. ed., D. Reidel, Dordrecht.
\bibitem{Char_2002}C.A. Charalambides, Enumerative Combinatorics, Chapman \& Hall / CRC, Boca Raton.
\bibitem{Jaynes_1957} E.T. Jaynes, {Phys. Rev.} 106 (1957) 620.
\bibitem{Johnson_K_1977}N.L. Johnson, S. Kotz, Urn Models and Their Application, John Wiley, NY, \S1.3.4.
\bibitem{Fang_1985}K.-T. Fang, {\it in} S. Kotz, N.L. Johnson (eds), {Encyclopedia of Statistical Sciences}, vol. 6, John Wiley, NY, 1985, 402-406.
\bibitem{Zwillinger_2003} D. Zwillinger, {CRC Standard Mathematical Tables and Formulae}, Chapman \& Hall / CRC Press, Boca Raton, FL, 2003.
\bibitem{Jordan_1947} C. Jordan, Calculus of Finite Differences, Chesea Publ., NY, 1947.

\end{thebibliography}

\end{document}